\documentclass[amsmath,amssymb,aps,twocolumn]{revtex4-2}
\usepackage{graphicx}% Include figure files
\usepackage{subfigure}
\usepackage{bm}% bold math
\begin{document}

\title{Quantum Superposition and Entanglement in Spin-Glass Systems}

\author{Asl{\i} Tuncer}
    \email[Correspondence email address: ]{asozdemir@ku.edu.tr}
    \affiliation{Ko\c{c} University, Institute of Physics, Sar{\i}yer, 34450, \. Istanbul,T\"urkiye}
 \author{Serhat C. Kad{\i}o\u{g}lu}   
    \affiliation{Ko\c{c} University, Institute of Physics, Sar{\i}yer, 34450, \. Istanbul,T\"urkiye}

\date{\today} 

\begin{abstract}
 We propose that spin glasses can exist in equally probable superposition states (SSs) comprising potential configurations. Employing the Edward-Anderson (EA) order parameter and magnetization, we establish a classification scheme for these SSs based on their contribution to discerning magnetic order (disorder), such as SG, ferromagnetic (FM), and paramagnetic (PM) phases. We also encompass various system sizes and investigate the entanglement properties of these phase-dependent SSs using the negativity measure. Our analysis reveals that the SG order parameter can be employed to determine the entanglement characteristics of magnetically ordered (disordered) phases, and vice versa, with negativity indicating the presence of magnetic order. Specifically, we examine the relationship between negativity and susceptibility in spin-glass systems. Our findings provide further insight into the role of quantum superposition in spin glasses and quantum magnets. 
 
\end{abstract}

\maketitle

\section{Introduction} \label{sec:intro}
Spin glasses are a fascinating phenomenon in condensed matter physics due to their unique microscopic properties. However, these systems contain random disorder, resulting in many possible states occurring with similar probabilities~\cite{FH1991,Mydosh,nish} and unable to arrange in a particular spin state, which satisfies the energy minimum for each interaction~\cite{Toulouse77II, Harary}. Such a situation is called frustration~\cite{Toulouse77I}. Even if a system is unfrustrated classically, it may exhibit frustration in the quantum case~\cite{illum2011,wolf2003,daw04,giampaola10,beaudrap10} due to non-commutativity and entanglement~\cite{daw04,adhikari2009}. Interestingly, even with just a few entangled elements, novel phenomena can occur in the quantum domain~\cite{daw04,adhikari2009, Parisi, Lewenstein}. Quantum fluctuations and entanglement can also play essential roles in the behavior of spin glasses since the spin-glass order occurs at low temperatures~\cite{Parisi}, and thermal fluctuations do not dominate the feature of spin-glass order. Quantum interference may also lead to unexpected effects, such as the suppression of tunneling \cite{Razavy} or the formation of localized states \cite{Anderson, MORO2018}. 

Spin glasses, on the other hand, should have frustrated spin(s). In recent years, many researchers have worked on these systems using spin-glass systems in the quantum domain~\cite{QannSG,harris2018phase, qsgslowdyn,nature2010,qsgdyn,KohKwek}. Although many properties of quantum spin glasses are examined, there is no direct approach to these systems over the superposition states, including the inherent frustration. 
We have investigated the presence of spin glasses within various quantum superposition states of potential electronic configurations, without the necessity of utilizing an ensemble of spin configurations or performing averages over replicas. Depending on the fact that the spin glasses exhibit a state of freezing within any of these possible configurations, we propose that these configurations can be defined through the superposition of computational basis vectors. We directly measured the local magnetic moments of the spins and the Edwards-Anderson SG order parameter to describe the corresponding magnetic phases, which include all-to-all interactions and randomly distributed antiferromagnetic(AFM)impurities. 

While previous studies have explored the entanglement properties of magnetic systems~\cite{Amico2009,Bennett1993} and also it was
shown that entanglement witnesses in spin systems can be related to thermodynamic quantities
~\cite{Toth2005,Wu2005,Hide2007}, there has been a lack of investigation into the direct relationship between entanglement and the spin-glass order parameter, as well as the absence of explicit characterization of single quantum states such as equally probably superposition states (Sec.~\ref{sec:model}).

In addition, quantum phase transitions (QPTs)~\cite{Sachdev,qptscience,Lewenstein,varsgSK} are studied in several methods for driving quantum states to a target state have been studied extensively in the literature. These include quantum entanglement, state transfer~\cite{stransf,stransf0,stransf1,stransf2,stransf3,stransf4}, and quantum adiabatic evolution~\cite{Joye1994,stransf5}. Given that we have obtained the corresponding quantum states for different magnetic phases, the study of phase transitions becomes feasible within this framework. We have investigated the occurrence of quantum phase transitions between these distinct magnetic phases, which is contingent upon the number of frustrated spins.

The rest of the following paper is organized as in Sec.~\ref{sec:model}; we introduce our model and describe the procedure for identifying the superposition states contributing to the SG phase. We also expand our results to well-known magnetic orders (disorder) and classify these superposition states (SSs) concerning their phase contributions, such as paramagnetic (PM) and ferromagnetic (FM) phases in Sec.~\ref{sec:develop1}. Once we identify the phase-based superposition states, we discuss the role of entanglement and the relationship between the SG-order parameter and negativity in Sec.~\ref{sec:entnglmnt}. Furthermore, we have conducted an inquiry into the direct correlation linking the zero-field susceptibility and negativity, specifically in relation to the spin-glass order parameter. We have reported that the quantum phase transitions can be detected depending on the frustrated spin number. Finally, we conclude the paper in Sec.~\ref{sec:conclusions} with an outlook on our results and the impact on future theoretical investigations and experimental implementations of current quantum technologies.
%%%%%%%%%%%%%%%%%%%%%%%%%%%%%%%%%%%%%%%%%%%%%%
\section{Model}
\label{sec:model}
%%%%%%%%%%%%%%%%%%%%%%%%%%%%%%%%%%%%%%%%%%%%%%
We consider $N$ quantum spins interacting through infinite-ranged exchange interaction with the Hamiltonian~\cite{ea75,Bray1980},
\begin{equation}
    \hat{H}=-\Sigma_{(i,j)} J_{ij} \hat{\sigma_i^z}\hat{\sigma_j^z}.
    \label{eq1}  
\end{equation}
Here $\hat{\sigma_j^z}$ is Pauli z-operators with $ I, j=1,\dots,N$, and the interaction couplings are quenched variables governed by a Gaussian probability distribution,
$P(J_{ij})=(2\pi N/J^2)^{-1/2} \exp {\frac{1}{2}(\frac{NJ_{ij}^2}{J^2})}$, with a variance $J^2/N$ and zero mean $<J_{ij}>=0$. The randomly distributed antiferromagnetic interaction is the source of frustration in case the system's Hamiltonian can not be minimized, at least for one spin or bound. Although there is no direct analogy between the geometric frustration in classical systems and its quantum counterpart, it has been defined in quantum systems related to entanglement and coherence effects~\cite{illum}. 

\begin{figure}
    \centering
    \includegraphics[scale=.36]{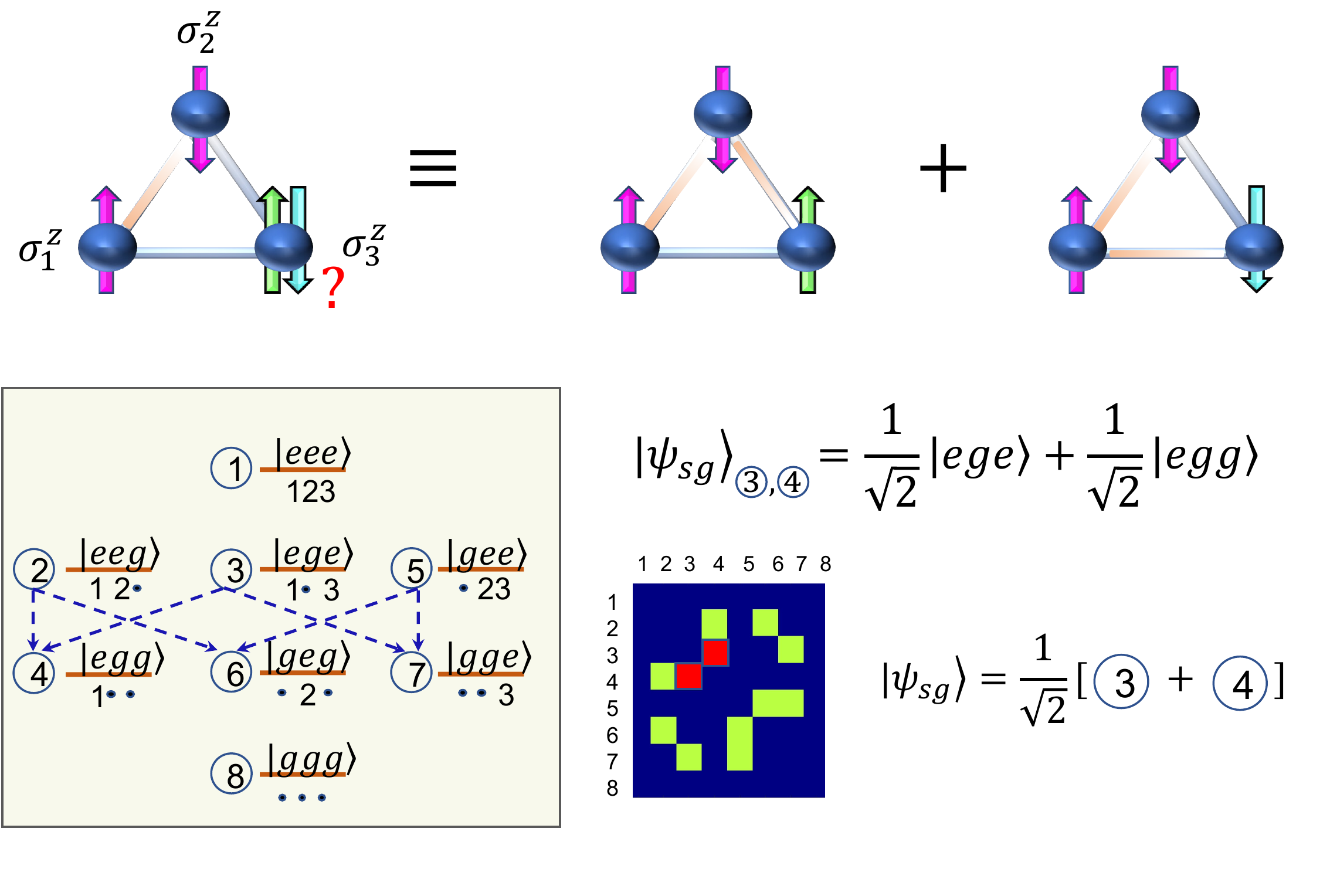}
    \caption{(Color online) {\it Top panel}: The graphical representation of SG-superposition state (SG-SSs) for $N=3$ qubits. The interactions (lines) between the qubits (blue spheres) are illustrated for FM $(J>0)$ and AFM $(J<0)$ couplings are yellow and light blue edges, respectively. {\it Bottom left}: The possible SSs are found by matching the dashed lines. {\it Bottom right}: One of the $|\psi_{SG}\rangle$ state contributes to the SG-order shown as red elements of the matrix representation. The numbers below the levels correspond to the labels of the excited spins of the state, and the dots represent the placement of the ground-level. The circled numbers next to the levels show the corresponding computational basis states. In the smaller box, the corresponding SG-contributed SSs of the three-spin system with $m=0$ and non-zero SG-order parameters are shown in green color.}
    \label{fig:1}
\end{figure}

We primarily focus on the $N$-spin system characterized by all-to-all interactions have spin permutation symmetry. Due to the binary nature of each spin, the phase space is comprising $2^N$ configurations. We commence with the simplest model of a three-qubit system with interactions, which may exhibit frustration, as depicted in Fig.~\ref{fig:1}. While thermal or quantum fluctuations drive the system towards a transition, this study does not account for thermal fluctuations.

To inject the quantum effects into the classical version of the Ising model, so-called the Edward-Anderson spin-glass Hamiltonian~(\ref{eq1}), the well-known ways is to add a transverse field. The quantum fluctuations arise from a competition between the spin-spin interactions and such an applied transverse external field. In contrast, in the present study, we propose that quantum fluctuations and frustration are introduced into the system via the direct injection of quantum superposition states. 

We were able to determine the specific superposition states that contribute to various magnetic phases through calculating the corresponding order parameters, and also categorizing states according to their magnetic properties, utilizing physical order parameters and entanglement, is a critical prerequisite for effectively manipulating entangled states necessary for quantum information processing and transfer via qubits~\cite{Bennett1993}. 

The EA spin-glass order parameter, which corresponds to overlaps of the local magnetization~\cite{ea75} and is provided in (\ref{eq2}), was also utilized in our analysis.
\begin{equation}
    q_{\text{EA}}^\alpha=\frac{1}{N}\sum_{i=1}^N (m_\text{i}^\alpha)^2.
    \label{eq2}  
\end{equation}     
In our case, the average magnetization becomes
\begin{equation}
    m=\frac{1}{N}\sum_{i=1}^N \langle \psi_{\text{suppos.}}|\hat{\sigma_i^z}|\psi_{\text{suppos.}}\rangle,
    \label{eq3}
\end{equation}
and the spin-glass order parameter includes the overlap between the SSs. In this way, we have found contributions to SG-order from the equally weighted superposition state in both energy and computational basis with the non-zero $q_{\text{EA}}$ and $m=0$. 
Besides obtaining the SSs contributing to the SG-phase, we obtained the PM-SSs in both vanishing $q_{\text{EA}}=0$ and $m=0$ and FM-SSs with $q_{\text{EA}}\neq0$ and $m\neq0$  cases~\cite{Sherrington1998}. The symmetry on the positive and negative magnetization values of the FM-SSs signify the spontaneous symmetry breaking in FM phase~\cite{spontsymBreak}. All these different order and disorder SSs are given in Sec.~\ref{sec:develop1}, and we explained explicit classification and association with their entanglement properties in Sec.~\ref{sec:entnglmnt}.

The state of a system consisting of $N$ spins can be considered a product state. The initial state is set to be a superposition of all possible configurations, $|\psi\rangle=\otimes_i^N |\rightarrow\rangle$ with $|\rightarrow\rangle=\frac{1}{\sqrt{2}}(|\uparrow\rangle+|\downarrow\rangle)$, where $|\uparrow\rangle$ and $|\downarrow\rangle$ are the eigenbasis of the Pauli-$z$ operator. However, the superposition states we have created by taking the sum of these product states will no longer be product states with the quantum correlations they contain. While the random disorder interactions between spins force a fixed orientation to minimize the energy, some spins may remain in the $|\rightarrow\rangle$ state even if there is no external field due to frustration. Thus the spins have three orders of freedom, such as $|e\rangle\equiv|\uparrow\rangle$, $|g\rangle\equiv|\downarrow\rangle$ and $|C\rangle\equiv|\rightarrow\rangle$, the resulting system state will be combinations of them, and the total number of configurations is $3^N$. We showed that the number of spins remaining in $|C\rangle$ directly relates to the system state's entanglement and order/disorder properties. 

In general, the qubit state can be in any superposition of the form $|\phi\rangle=a_1|0\rangle+a_2|1\rangle$
as long as $|a_1|^2+|a_2|^2=1$. However, this is not the case for classical spins. As the state $|C\rangle$ is the special case of the of $a_1=a_2=1/\sqrt{2}$ such as called \textit{cat state}, we illustrate such a frustrated configuration state of a three-body system, $|\psi>=|0\rangle\otimes|1\rangle\otimes|C\rangle$, in Fig.~\ref{fig:1} separated into two states that do not have frustration in the $\sigma^z$ basis:
\begin{equation}
\begin{aligned}  |\psi>=\frac{1}{\sqrt{2}}\left (|0\rangle\otimes|1\rangle\otimes|0\rangle+|0\rangle\otimes|1\rangle\otimes|1\rangle \right )
\end{aligned}
\label{eq5}  
\end{equation}  
where $|0\rangle\equiv|e\rangle$ and $|1\rangle\equiv|g\rangle$ in the standard basis of the energy-state. This standard basis is ${|e\rangle=(1~~0)^\dagger, |g\rangle}=(0~~1)^\dagger$ so called excited and ground-state for one atom, ${|ee\rangle, |eg\rangle, |ge\rangle, |gg\rangle}$ for two, and ${|eee\rangle, |eeg\rangle, |ege\rangle, |gee\rangle, |egg\rangle, |geg\rangle, |gge\rangle, |ggg\rangle}$ for three atoms. We will continue with the standard basis from now on. Fig.~\ref{fig:1} shows the three-atom standard basis on the bottom left. Corresponding positions in the natural basis of the levels are given in the circles next to levels, and the numbers with the green dots below the levels denote the placement of the excited spins and ground state, respectively. 

By considering the binary superposition states, all our states contribute to the spin-glass order should satisfy the two simple rules:
\begin{itemize}
    \item[(i)] The SG-SSs must have at least one co-excited spin in its superposed ones, 
    \item[(ii)] The total number of excited spin labels should not exceed the total number of spins in the system.
\end{itemize}

The first rule refers to the overlapping between the states, while the second one corresponds to the SS should have at least one frustrated spin an equally-weighted qubit-state, $|C\rangle$. Once we address the excited spins, we can write the spin-glass SSs easily in each system size. However, in FM SSs, the number of excited labels can be greater than the number of spins. Understandably, the ferromagnetic order does not have to include any frustrated spin, and all the spins can be in the same (up or down) direction. So the self-superposition of all spins up (down) will be one of the solutions of the FM system state. Fixing a spin on a state can be thought of as making a local measurement on it, and this causes a loss of entanglement. We examined this loss in terms of negativity, a measure of quantum entanglement defined as~\cite{Vidal},
\begin{equation} 
\mathcal{N}=\frac{\lVert \rho^\mathrm{T} \rVert_1-1}{2}. 
\label{eq6}
\end{equation}
Here, the trace norm of the quantum state $\rho=|\psi_{\text{suppos.}}\rangle \langle\psi_{\text{suppos.}}|$ is denoted by $\lVert \rho^\mathrm{T} \rVert_1$. The details of numerical calculations and negativity results are explained in Sec.~\ref{sec:entnglmnt}.
%%%%%%%%%%%%%%%%%%%%%%%%%%%%%%%%%%%%%%%%%
\section{Quantum Superposition States of Spin-glass, FM, and PM phase} \label{sec:develop1}
%%%%%%%%%%%%%%%%%%%%%%%%%%%%%%%%%%%%%%%%%
The magnetic phases of interest can be classified in terms of magnetization, and the EA spin-glass order parameter,  $q_{\text{EA}}$~\cite{ea75}. 
\begin{figure}[ht]
\centering   
    \subfigure[Order parameters for all superposition states at $\text{N}=3$.]{
    \includegraphics[width=.8\linewidth]{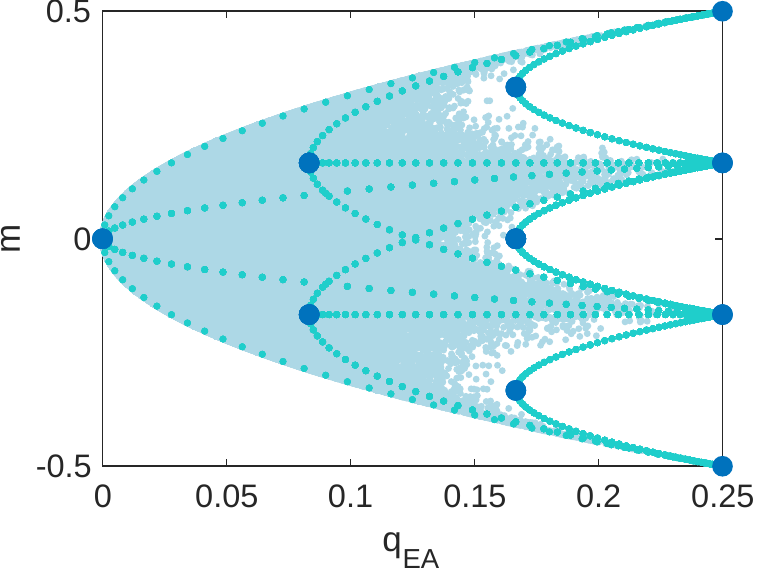}   
    \label{fig2a}
    }

    \subfigure[Order parameter for equally weighted binary SSs.]{
    \includegraphics[width=.85\linewidth]{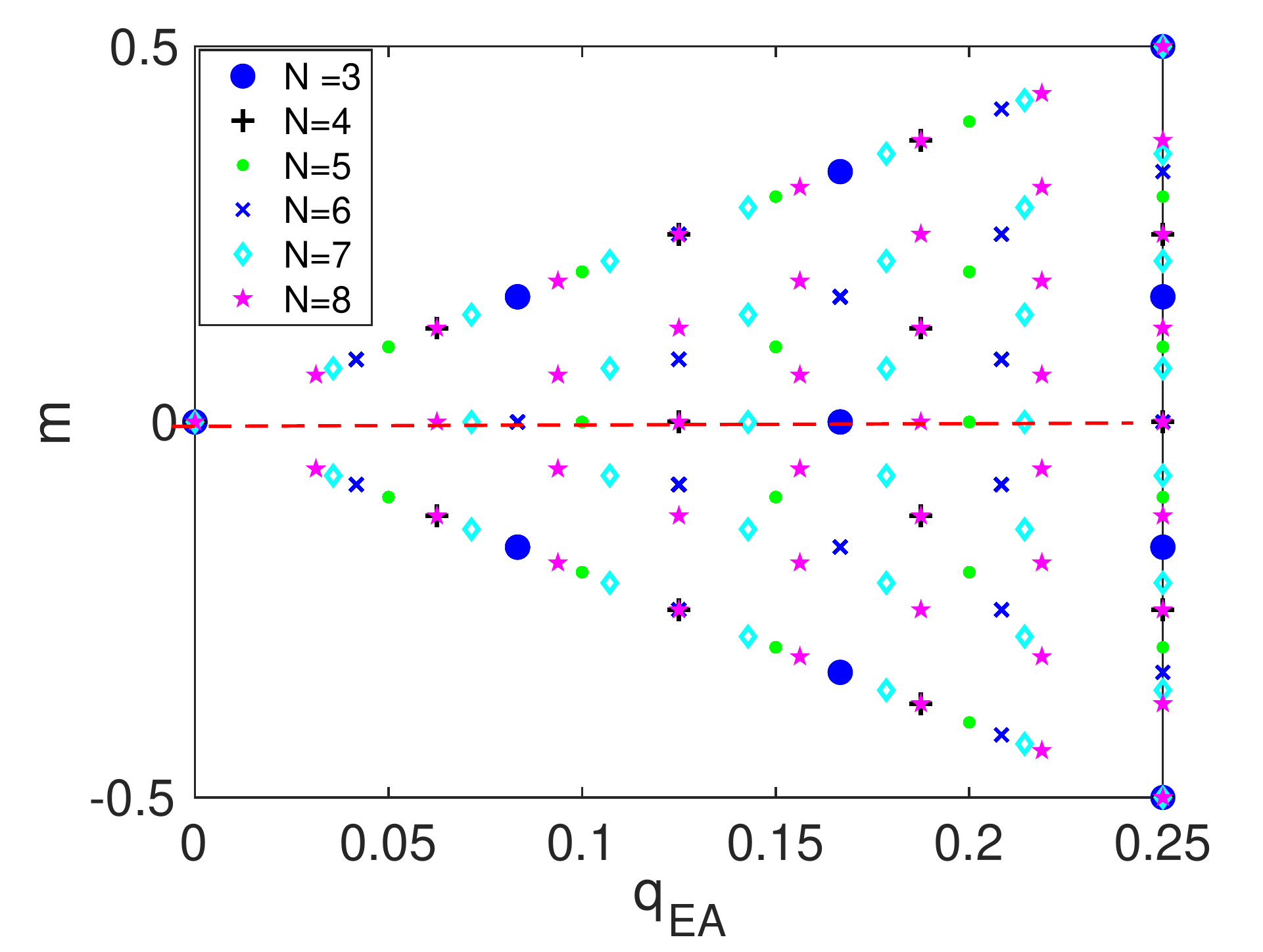}
    \label{fig2b}
    }
\caption{(Color online.) The magnetization vs. EA spin-glass order parameter, $q_{\text{EA}}$, is shown for (a) randomly weighted all SSs at $N=3$. While the dotted lines denote all binary SSs, dark blue circles show the equally probable binary SSs, (b) all equally-weighted binary SSs in different sizes from $N=3,\dots,8$. The red horizontal line indicates $q_{\text{EA}}\neq 0$ with $m=0$ corresponding to the SG regime. The PM phase is observed in both figures at $m=0$ and $q_{\text{EA}}=0$.}
\label{fig:2}
\end{figure} 
Once we need to define the system's state to calculate these order parameters. 
\begin{figure*}[ht]
    \includegraphics[scale=.3]{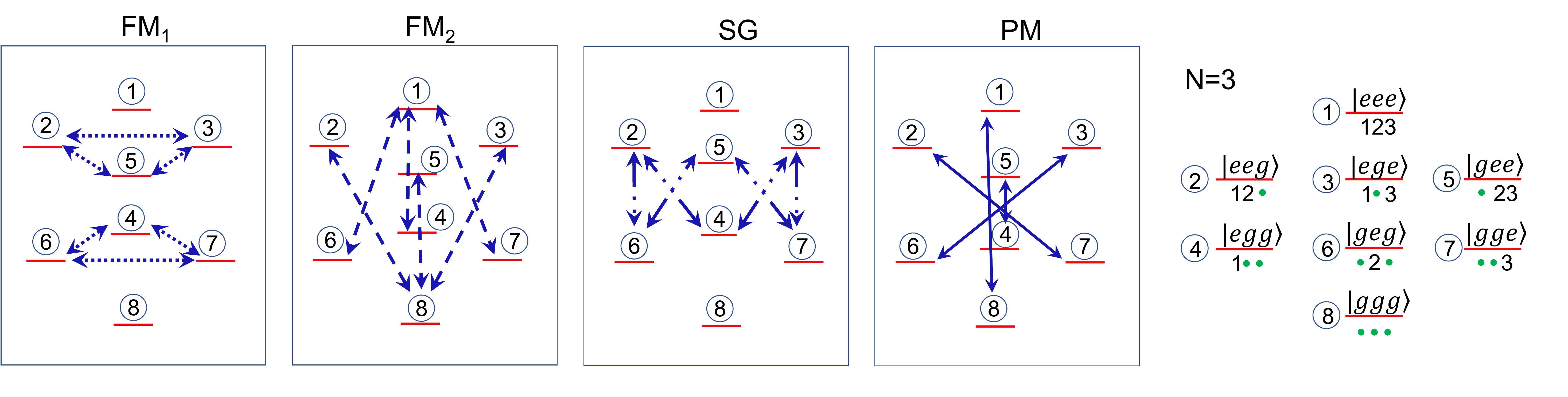}
    \caption{(Color online.) All equally likely binary SSs and their corresponding magnetic orders at $N=3$ qubits are illustrated. The first two boxes contribute to ferromagnetic phases with positive and negative average magnetization, $\text{FM}_1$ and $\text{FM}_2$, respectively. The third and the last boxes display SG and PM phase contributions. All arrows depict the equally weighted summation of the states.}
    \label{fig:3}
\end{figure*}

\indent \textit{Randomly weighted superposition states}. In general, the SSs can be taken as randomly weighted superposition states of $N$-spin system is 
\begin{equation}
|\psi_{\text{suppos.}}\rangle=\sum_{i=1}^{2^N}\alpha_i|\psi_i\rangle,
    \label{eq:7}
\end{equation}
here $i=1,\dots,2^N$ is the number of the basis vectors with random complex numbers, $\alpha_i$. We showed the domain of the corresponding order parameter values at $N=3$ for randomly superposed $5$ million states' space in Fig.~\ref{fig2a}. The dotted lines inside the main parameter domain denote all binary superposition states with and without equal probable ones. In addition, we specifically indicated and marked the equally likely superposition states with dark blue circles. Even though we confine the system states to the subset of equally weighted superposition space, which constitutes a fraction of the entire set of states, we observe that we consider both a region close to the boundaries of the parameter space and states that contribute to distinct phases. 

\indent \textit{Equally likely binary SSs}. By considering only equally likely binary superposition states in the computational base, obtaining a set of states that contribute to the FM, PM, and SG phases provides us with significant convenience in both quantum computation and experimental studies on spin glasses~\cite{qsgdyn,nature2010,qsgslowdyn}. These binary superposition states can be written as
\begin{equation}
|\psi_{\text{suppos.}}\rangle=\alpha|\psi_1\rangle+\beta|\psi_2\rangle,
    \label{eq:8}
\end{equation}
here $\alpha, \beta$ are the complex numbers, and $|\psi_1\rangle, |\psi_2\rangle$ are the superposed states which can either be the eigenstates of the system Hamiltonian or computational basis states. Note that, from the conservation of probability, $|\alpha|^2+|\beta|^2=1$ is likely to have infinitely many different SSs. Therefore, one can expect to have infinitely many order parameter values by substituting these SSs into~(\ref{eq2}) and~(\ref{eq3}). However, we have shown the confined region of the order parameter values in Fig.~\ref{fig:2}. Though the region depends on the choice of the SSs, the certain triangular domain is illustrated in Fig.~\ref{fig2b} for equally weighted binary superposition states at $\alpha=\beta=1/\sqrt{2}$ in~(\ref{eq:8}) even if we go to the larger system sizes. 

From now on, we will continue with the equally likely binary SSs and note that we could only reach the states contributing to the spin-glass phase if we took the values of $\alpha$ and $\beta$ with equal weights as $1/\sqrt{2}$ in~(\ref{eq:8}). It is understandable to consider the frustration condition in spin-glass systems discussed above. Adding a phase, represented by $\alpha=\exp(-i\gamma)$, to the superposed states does not affect the values of the order parameters. This outcome, which holds for any real number $\gamma$, can be readily derived as a straightforward result. Figure~\ref{fig:3} shows equally weighted binary superpositions in size $N=3$ correspond to which different quantum magnets, including SG, FM, and PM-phases. 

\indent \textit{Spontanous symmetry breaking}. In both cases of Fig.~(\ref{fig:2}), having various numbers of $m$ values against the same $q_{\text{EA}}$ value indicates spontaneous symmetry breaking~\cite{spontsymBreak}. While this symmetry breaking signifies the FM order, the PM phase corresponds to the point of $m=0$ and $q_{\text{EA}}=0$. The red-dashed line starts with the PM regime, and along the line, we can obtain the SSs corresponding to the SG regime. The quantum phase transition can be considered from the PM to the SG phase. 
\begin{figure}[h]
\centering
\includegraphics[scale=.23]{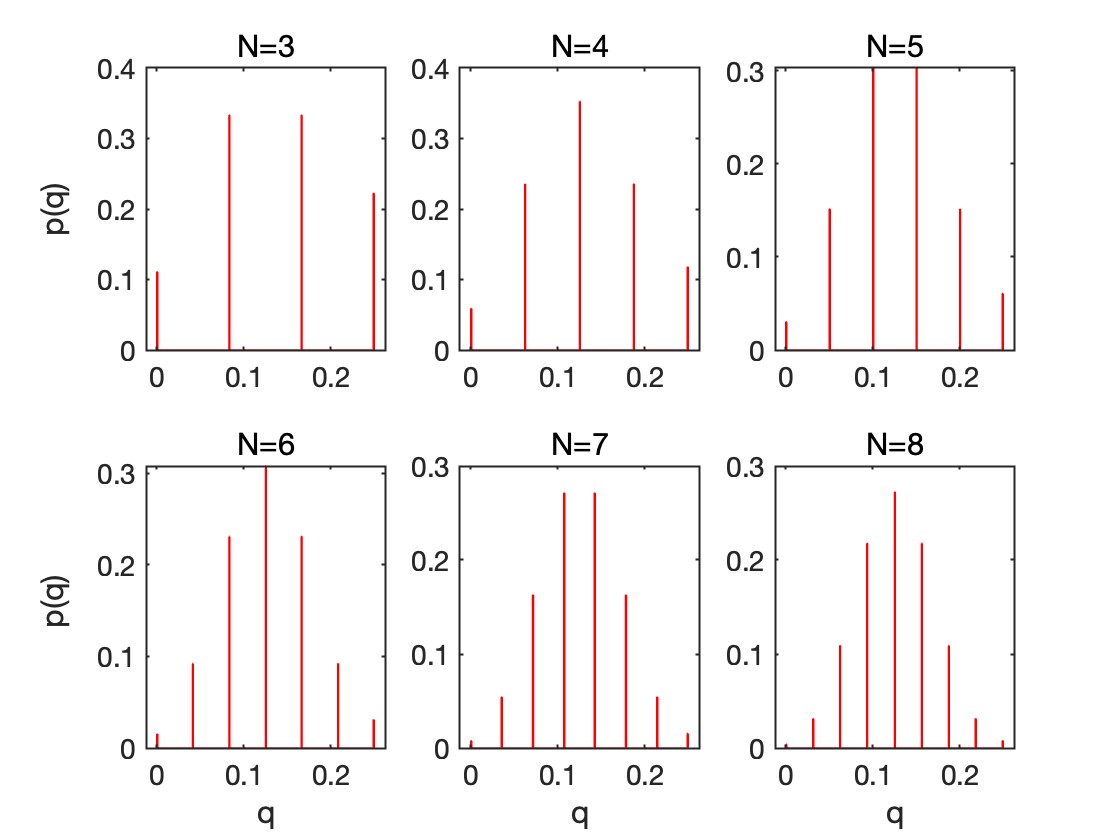}
\caption{(Color online.) EA spin-glass order parameter probability distribution is given for different system sizes.}
\label{fig:4}
\end{figure}
Even if the system size enlarges and extra SSs arise, the triangular structure and the boundaries of the diagram remain the same. In other words, the symmetry breaking can be observed for $N\to \infty$ in the different $q_{\text{EA}}$ values. Moreover, the Edward-Anderson spin-glass order parameter can not surpass the maximum value of $q_\text{EA}=0.25$ as independent of the choice of SSs. 

In Figure~\ref{fig:4a}, the matrix-like representation of the binary SSs that contribute exclusively to the SG order in systems comprising $N=3,4,5,$ and $6$ qubits is given in the first row and the explicit depiction of the SSs of other magnetic phases in the second row of the figure. Although the contribution of SSs to the SG and FM phases varies as the system size changes, non-diagonal elements consistently counteract the PM phase due to their formation from non-overlapping product spin states. 

Furthermore, the upper and lower triangular regions of the non-diagonal elements in the matrix are associated with the FM phase depending on the specific binary superposition states involved. Consequently, this superposition matrix can be regarded as an evenly distributed superposition state space that pertains to the configuration space. The matrix is divided into distinct quantum magnetic phases, serving as a phase diagram. Additionally, we have noted that the phase partition pattern remains consistent and scalable even as the system size is expanded.
\newline
\indent \textit{Replica symmetry breaking}. The differentiation of the $q_{\text{EA}}$ order parameter at each even value of system size, $N$, can be associated with replica symmetry breaking in spin-glass systems~\cite{Parisi, Bray1980}. For instance, while there is a unique $q_{\text{EA}}$ value for $N=3$, sizes $N=4$ and $N=5$ have two different values of $q_{\text{EA}}$. 

Moreover, Fig.~\ref{fig:6} illustrates a recursive pattern, wherein taking a partial trace over the most recently added qubit leads the system state space to the previous state space. Co-excited atoms in SSs can also be defined as overlapping between two states. This overlap scales with the system size, similar to the differentiation of $q_{\text{EA}}$. The number of possible overlapping increases with each even number of $N$. However, unlike the SG-SSs, the PM-SSs have neither co-excited spins nor overlap. All PM-SSs are maximally entangled, similar to Greenberger-Horne-Zeilinger (GHZ) states~\cite{ghz}, specifically for binary superposition states. We illustrated that the entanglement of spin-glass SSs in the first line of Fig.~\ref{fig:4b} and the entanglement of SSs correspond to the other magnetic phases in the second line, respectively.
\begin{figure}[ht]
    \includegraphics[scale=.44]{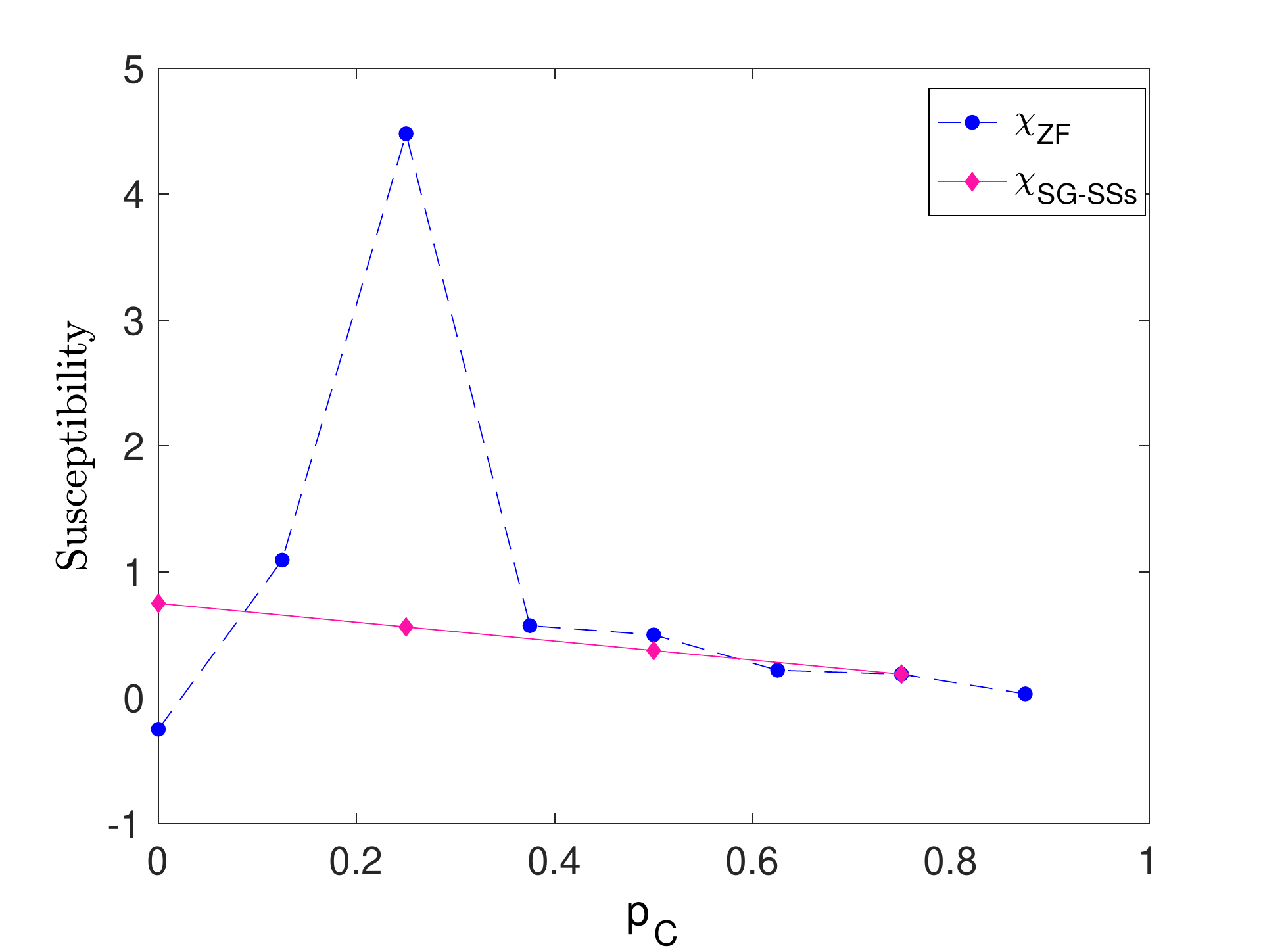}
    \caption{(Color online.) All equally weighted binary SSs and their corresponding magnetic orders at $N=8$ qubits are illustrated. }
    \label{fig:5}
\end{figure}
\begin{figure*}[ht]
\centering   {
    \subfigure[The non-zero SG order parameter $q_{\text{EA}}$ of equally weighted SSs.]{
    \includegraphics[width=.98\linewidth]{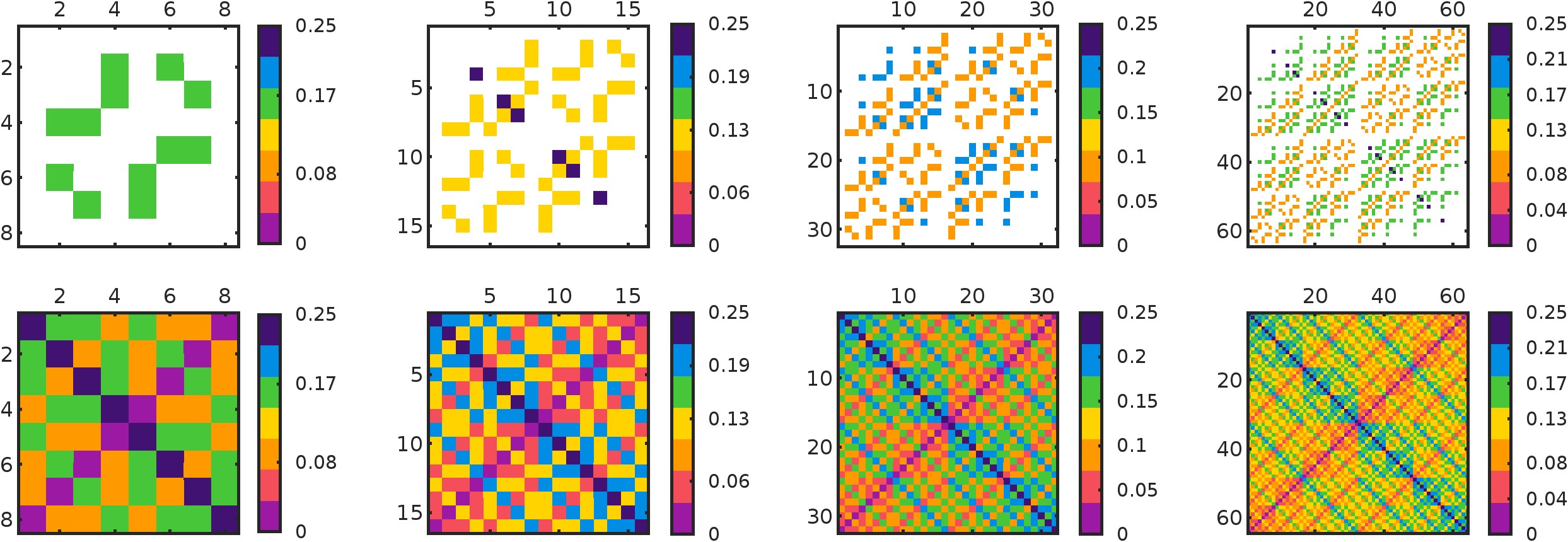}   
    \label{fig:4a}
    }

    \subfigure[First line: The Entanglement of equally weighted SSs contribute to SG order. Second line: Entanglement of all magnetic orders.]{
    \includegraphics[width=.98\linewidth]{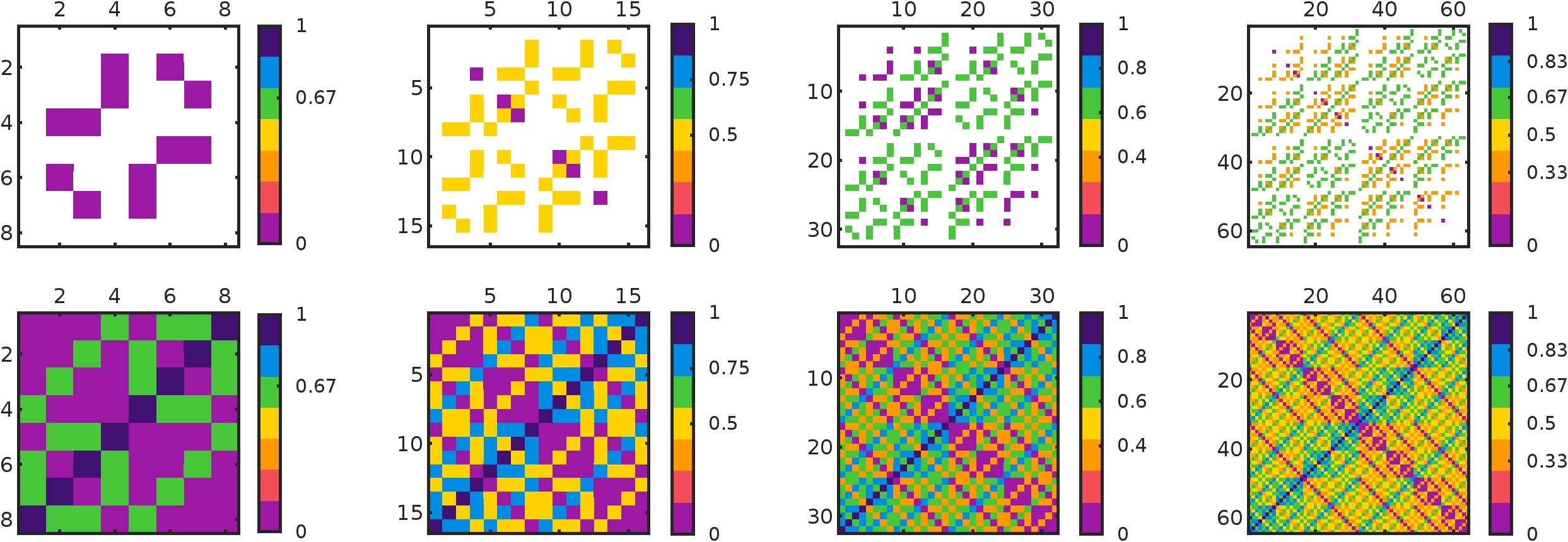}
    \label{fig:4b}
    }
    }
\caption{(Color online.) (a) The matrix-like representation of all SG order contributed equally likely SSs is given for $N=3,4,5$, and $6$ spins from left to right with non-zero EA spin-glass order parameter and zero magnetization. As the system size increases, the number of distinct values of $q_{\text{EA}}$ increases, indicating a signal of replica symmetry breaking~\cite{Bray1980, Parisi}. The various colors denote different $q_{\text{EA}}$ (or $\mathcal{N}$) values, and this pattern repeats itself in subsequent generations from left to right. Diagonal elements correspond to the single states, equal to $q_\text{EA}=0.25$ maximum value. However, their contribution to spin-glass or FM orders may alter by the system size. Another fixed part of the matrix representation is that off-diagonal elements have only $q_{\text{EA}}=0$ values and contribute to the PM order for all system sizes. (b) The negativity between each spin and the remaining part of the system is computed by averaging over all spins. While the same pattern of $q_{\text{EA}}$ obtained for average negativity $\mathcal{N}$, there is a reciprocal relationship between the $\mathcal{N}$ average negativity and the $q_{\text{EA}}$ order parameter.}
\label{fig:6}
\end{figure*}

\indent \textit{Zero-field susceptibility}. Before discussing the entanglement structure of these SSs, one of the most important quantity should be considered is the zero-field susceptibility defined in terms of the magnetic fluctuations ~\cite{magsuss,susfid},
\begin{equation}
    \chi=\tilde\beta(\langle m^2 \rangle-\langle m \rangle^2),
    \label{eq:b8}
\end{equation}
here $\tilde\beta\propto1/T$ inverse temperature. The average over all magnetic moments is $\langle m\rangle$, and $\langle m\rangle^2\equiv q_{\text{EA}}$ while $\langle m^2\rangle=1$ for Ising spin-glass in the thermodynamic limit, $N\to\infty$. We obtain the zero-field susceptibility so-called linear response susceptibility~\cite{mezard} by substituting $\langle m\rangle^2$ and $\langle m^2\rangle$ into~(\ref{eq:b8}),
\begin{equation}
    \chi=\tilde\beta(1-q_{\text{EA}}).
    \label{eq11new}
\end{equation}
Explicit derivation of the EA spin-glass order parameter for equally probably superposition states is given in Section~\ref{sec:apx} as $q_{\text{EA}}=1/4(1-p_C)$ in~(\ref{eq:b7}). By substituting (\ref{eq:b7}) into (\ref{eq11new}), the superposition-state zero-field susceptibility becomes
\begin{equation}
    \chi_{SS}=\tilde\beta(1-\frac{1}{4}(1-p_C)),
    \label{eq20}
\end{equation}
and for PM phase susceptibility, the probability of the superposed spins is $p_C=1$, will be $\chi_{SS}\propto\tilde\beta$ as convenient to the Curie law\cite{curie}.
However, we can modify this expression for the finite system size. Due to the local magnetic moments can take only the values ($1,-1,0$) of ($e,g,C$), the Equation ~(\ref{eq3}) depends on the number of $e$, $g$ and $C$ spins in SS states. Assuming the number of $C$-spins is $n_C$, only contribution comes from the remaining part, then the square average of the local magnetic moments is
\begin{equation}
    \langle m^2\rangle=\frac{N-n_C}{N}=1-\frac{n_C}{N},
    \label{eq12new}
\end{equation}
and the linear response susceptibility for the equally weighted superposition states will be 
\begin{equation}
   \chi_{suppos} =\tilde\beta[(1-p_C)-q_{\text{EA}}]
    \label{eq13new}    
\end{equation}
in terms of the probability of the number of the superposed spins ($C$), $p_C\equiv n_C/N$, and EA SG-order parameter. Here $(1-p_C), $ replaces $1$ in~(\ref{eq11new}) since the classical Ising spins do not become in a $C$ state, the $p_C$ probability is zero.
Figure~\ref{fig:5} presents the results obtained from Equation ~(\ref{eq13new}) by substituting the numerical values of $q_{\text{EA}}$ and the corresponding $p_C$ probabilities. The blue dashed line represents the average zero-field susceptibility over all superposed states (SSs), while the solid pink line specifically depicts the contribution of SSs in the SG order. When considering all SSs, the zero-field susceptibility ($\chi_{ZF}$) exhibits a cusp at $p_C=0.25$, indicating the presence of ferromagnetic (FM) order. Additionally, for $p_C>0.25$, the SSs are associated with the SG order. This segregation of SSs signifies a quantum phase transition from FM to SG.
 % &=\tilde\beta(4q_{\text{max}}-q_{\text{EA}}) 
%%%%%%%%%%%%%%%%%%%%%%%%%%%%%%%%%%%%%%%%%%%%%%%%%
\section{The New Entanglement Witness of Magnetic Structures} \label{sec:entnglmnt}
%%%%%%%%%%%%%%%%%%%%%%%%%%%%%%%%%%%%%%%%%%%%%%%%%

In order to obtain the entanglement value between each spin and the rest of the system, an average was taken over all spins.
\begin{figure*}[ht]
\centering
\includegraphics[scale=.4]{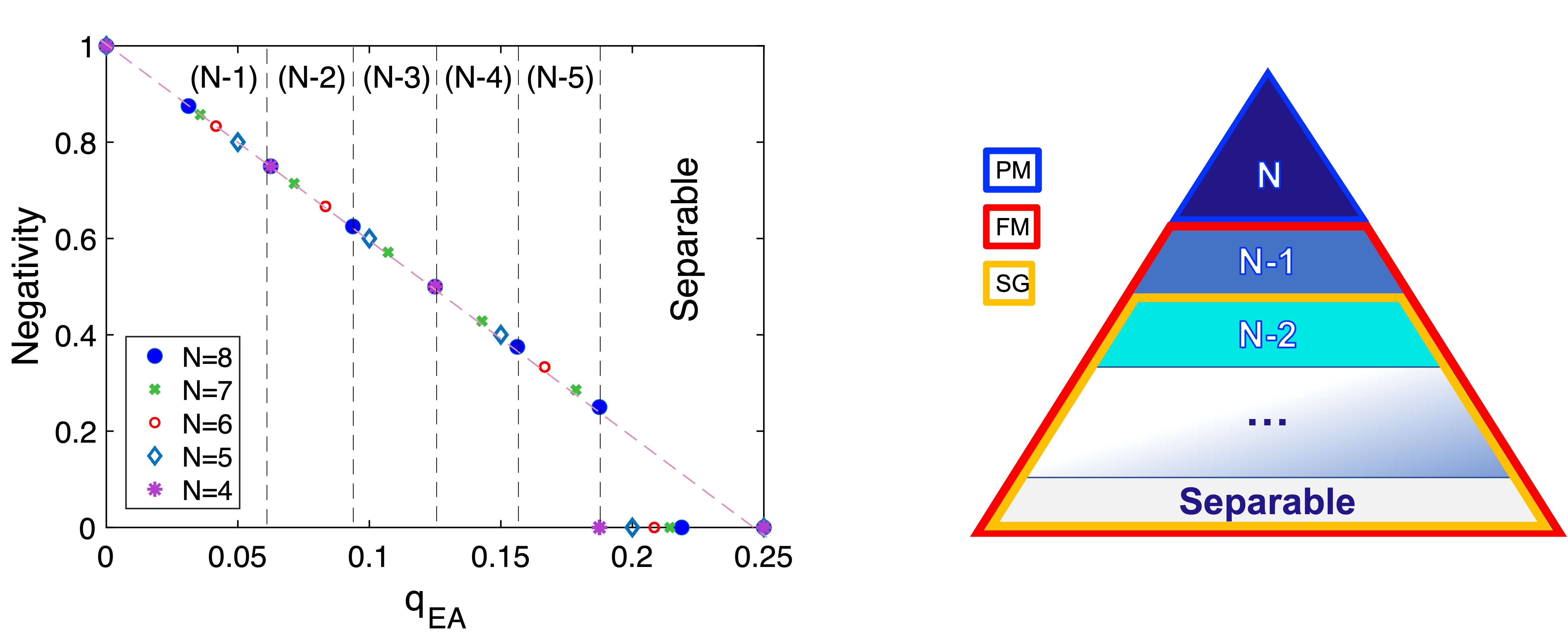}
\caption{(Color online.) The left panel shows the variation of the negativity with the Edward-Anderson spin-glass order parameter $q_{\text{EA}}$ for particle systems with $N=4,5,6,7$, and $8$. The distinct regions in the plot correspond to different values of $q_{\text{EA}}$ that indicate the extent of particle entanglement. Initially, all systems are fully entangled with $\mathcal{N}=1$, and as $q_{\text{EA}}$ decreases, the number of entangled particles decreases, ultimately leading to a separable state with $\mathcal{N}=0$. In the right panel, the phase-contributed superposition states are classified based on the number of entangled particles in the $N$-particle system.}
\label{fig:7}
\end{figure*}
The presence or absence of overlap between states in their superposed states corresponds to the ordered/disordered states of the system. Moreover, the overlapping superposition states can exhibit different magnetic orders, such as SG, FM, and antiferromagnetic orders. Firstly, after defining the order/disorder distinction based on the presence or absence of the overlap, we observe that PM (disordered) systems corresponding to zero overlaps also possess maximum entanglement. From this perspective, we calculated the entanglement of the SSs using negativity to define the relationship between the amount of overlap and entanglement. Considering that the EA spin-glass order parameter measures the degree of overlap between two different system configurations (superposed states), we investigated the direct relationship between negativity $\mathcal{N}$ and $q_{\text{EA}}$ order parameter. The numerical results illustrate the relationship between these two quantities, as shown in Fig.~\ref{fig:7}. 
We obtained the negativity decreases linearly with $q_{\text{EA}}$,
\begin{equation}
\mathcal{N}=1-\frac{q_{\text{EA}}}{q_{\text{max}}}.
\label{eq14}
\end{equation}

Here, $q_{\text{max}}=0.25$ is the maximum value of the EA spin-glass order parameter, and explicit derivation is given in Sec.~\ref{sec:apx}. 
It should be noted that, in the quantum droplet (cluster) model of Thill and Huse~\cite{droplet1,droplet2}, the relative reduction of the temperature-dependent SG order parameter from its zero temperature value is given as 
\begin{equation}
1-\frac{q_{\text{EA}}(T)}{q_{\text{EA}}(0)}\sim \frac{\hat{\beta}}{L(T)},
\label{eq15}
\end{equation}
here $L(T)$ is the length of the droplets in the system at enough low temperature~\cite{busiello}, and it is given as the classical-to-quantum crossover length scale~\cite{busiello,droplet1,droplet2}. We have shown that the quantity is equal to the average of the bipartite negativity of the superposition states as $q_{\text{max}}$ corresponding to the zero-temperature spin-glass order parameter $q_{\text{EA}}(0)$. 

Despite recent investigations into the entanglement characteristics of the Quantum Sherrington-Kirkpatrick Model (QSKM)~\cite{2022cirac} incorporating all-to-all interactions, there remains a dearth of literature addressing the explicit connection between the entangled portion and the order parameter associated with the Spin Glass (SG) phase or other magnetic phases. In Eq.~(\ref{eq14}), the relationship between negativity and spin-glass order parameter has yet to be documented. We obtained not only the correlation between the reduction of the EA spin-glass order parameter and the frustrated spin ratio~(\ref{eq:b7}) but also the dependence of the linear response susceptibility on the entangled portion of the system state~(\ref{eq14}). From this perspective, the zero-cooled-field susceptibility can be written regarding the average bipartite negativity~(\ref{eq:15}). 

As the negativity depends on the $q_{\text{EA}}$, and by considering Eq.~(\ref{eq14}) and~(\ref{eq20}) susceptibility can be written in terms of the negativity,
\begin{equation}
    \chi_{ZFC}=\frac{\tilde\beta}{4}(3+\mathcal{N}).
\label{eq:15}
\end{equation}

To investigate the relationship, we examine the negativity value for the paramagnetic case, which is $\mathcal{N}=1$. This yields $\chi_{ZFC}=\tilde\beta$, aligning with Curie's law~\cite{curie}.

Furthermore, the numerical calculations demonstrate consistent correlations between negativity and the spin-glass order parameter. According to the Fig.~\ref{fig:7}, the state starts from the fully entangled one with $\mathcal{N}=1$ and $q_{\text{EA}}=0$ (PM state). Then, its maximal-entangled portion becomes smaller and smaller until it reaches the separable state with $\mathcal{N}=0$ and $q_{\text{EA}}=q_{\text{max}}=0.25$. Each distinct $q_{\text{EA}}$ value corresponds to a different entangled cluster size. The linear dependence of negativity on the size of the entangled part of the system indicates the area law of the entanglement distribution~\cite{arealaw,mailent}.

We separated regions corresponding to $n$-partite entanglement by dashed lines with $n=N, N-1,\dots,1$. However, as the system size approaches infinity ($N\to \infty$), the classification mentioned above is no longer discernible, and the different $q_{\text{EA}}$ values on the fitting line be indistinguishable. The separations between distinct $q_{\text{EA}}$ values vanish due to the loss of quantumness in the system as it becomes classical.

In this analysis, we present both numerical results and a discussion of SSs, including the multi-partite entangled component, concerning the recursive growing pattern of the SG-order parameter $q_{\text{EA}}$ and negativity $\mathcal{N}$, illustrated in Fig.~\ref{fig:6}. Specifically, we consider a three-particle system where each particle can exist in one of the three configurations from the set $E_{N=3}(SG)=\{C,e,g\}$, and $E_{N=3}(SG)$ denotes the ensemble including the probable spin configurations satisfies the six possible configuration states for the SG contribution with vanishing magnetization
\noindent and non-zero $q_{\text{EA}}$.
Suppose one more qubit is added to the system. The system state has two possible configuration ensembles: 
\begin{equation}
E_{N=4}(SG) \\ 
=
\begin{cases}
   \{C,C,e,g\}, &q_{\text{EA}}=0.125, \\
   \{e,g,e,g\}, &q_{\text{EA}}=0.25. 
\end{cases}
\label{eq10}
\end{equation}
which corresponds to SG-contributed superposition. Two distinct $q_{\text{EA}}$ values correspond to the two different configuration sets. The permutation group of the first set yields 24 different SSs that have $q_{\text{EA}}=0.125$, while the second set yields six SSs with $q_{\text{EA}}=0.25$ appropriate to Fig.~\ref{fig:6}. As we increase the system size to $N=5$ spins, the number of possible sets remains the same as for $N=4$, and we observe two additional sets, namely $\{C,C,C,e,g\}$ and $\{C,e,g,e,g\}$. 

Notably, the $q_{\text{EA}}$ parameter differs between distinct sets and within a set, depending on the number of entangled particles. For instance, the set $\{C,C,C,e,g\}$ can be considered as 
\begin{equation}
\{C,C,C,e,g\} \\ 
=
\begin{cases}
   \{|GHZ\rangle_3, e, g \} \\
   \{|GHZ\rangle_2,C,e,g\},  
\end{cases}
\label{eq11}
\end{equation}
here, the subset $|GHZ\rangle_3:\{C,C,C\}$ and $|GHZ\rangle_2:\{C,C\}$ gives the maximally entangled (GHZ) states, and the subscript denotes the number of entangled particles. In general definition, $n$-particle GHZ state can be written as 
\begin{equation}
 |GHZ\rangle_n=\alpha(|g\rangle^{\otimes n}+|e\rangle^{\otimes n}),
\label{eq12}
\end{equation}
where $\alpha$ is a constant~\cite{ghz,nielsen}, the source of these maximally entangled states is related to the number of $C$ in the permutation sets.

If the possible configuration sets lack  $|C\rangle$, the resulting state will be deemed \textit{separable}. The superposition state may also be partially entangled, in which the number of cat states is less than $N-1$. Defining the entangled portion of the state, particularly for larger systems, presents a challenge. While our focus centers on SG superposition states, analogous conditions arise in the FM case. 

We developed a metric to quantify the entanglement partition of a state in terms of the spin-glass order parameter, $q_{\text{EA}}$, which allowed us to achieve our objective. 
Fig.~\ref{fig:7} illustrates the entanglement partitions for various system sizes and the inverse linear relationship between the negativity and $q_{\text{EA}}$. 
As $q_{\text{EA}}$ decreases from its maximum value corresponding to separable states, the number of entangled particles increases until the system reaches a state of maximum entanglement. 

This classification of the magnetic phases is based on the presence of entangled particle ensembles. From this aspect, it can be remarked as an analogy between locally ordered (disordered) parts of the whole system and locally entangled portions.  
\section{Conclusions} \label{sec:conclusions}
This research demonstrates that spin glasses can exist in equally likely superposition states of potential electronic configurations in a quantum framework—the inherent uncertainty or ambiguity associated with cat states in quantum mechanics. We propose using cat states to define frustrated spins and link the frustration to quantum interference. By employing the EA spin-glass order parameter and magnetization, we classify the superposition states based on their contribution to distinguishing magnetic order (or disorder) in various phases, such as SG, (anti)FM, and PM. Our results provide valuable insights into the nature of spin glasses in quantum systems and have implications for developing quantum technologies such as quantum cryptography~\cite{Pirandola}, quantum simulation~\cite{georgescu,daley2022,brown2010}, quantum computation~\cite{Bennett, Valiev,Kim,Normand}, quantum sensing~\cite{qusens} and metrology~\cite{Schaetz}.

We establish a direct correlation between the Edward-Anderson spin-glass order parameter, $q_{\text{EA}}$, and a measure of entanglement, negativity $\mathcal{N}$. We demonstrate that the spin-glass order parameter can also function as an indicator of entanglement. Since the FM and SG phases overlap concerning corresponding SSs' entangled partition, the negativity of entanglement can serve as the order parameter to distinguish between phases of order and disorder, specifically the ferromagnetic (FM) and paramagnetic (PM) phases. This result is because entanglement is the ability of qubits to correlate their state with other qubits.

This study reveals that the structural similarities between entanglement and spin-glass order parameters persist across systems of varying sizes, as indicated by the consistent patterns shown in Fig.~\ref{fig:6}. Furthermore, our study highlights the potential use of superposition states in defining magnetic order (disorder)~\cite{Schaetz} in condensed matter physics, which has broader implications for quantum information processing and quantum computing~\cite{Bennett, Valiev,Kim,Normand,Bennett1993}. These findings offer new insights into the role of quantum superposition in spin glasses and quantum magnets. 
We offer that these superposition states are candidates to be new phase-based bits in quantum computing~\cite{nielsen,Feynman}.

\section*{Acknowledgements} \label{sec:acknowledgements}
 The authors would like to acknowledge the financial support from the Scientific and Technological Research Council of T\"{u}rkiye (T\"{U}B{\. I}TAK), grant No. 120F100. We would also like to express our gratitude to Ö. E. Müstecapl{\i}o\u{g}lu for his insightful discussions.

\appendix

\section{Explicit calculation of EA order parameter over the binary-superposition states}\label{sec:apx}
Since we restrict ourselves to equally weighted binary-superposition states along the paper, we will start with the general binary superposition states~(\ref{eq:8}) given as a summation of the states $|\psi_1\rangle=|\alpha_1\alpha_2\dots\alpha_\text{N}\rangle $ and $|\psi_2\rangle=|\beta_1\beta_2\dots\beta_\text{N}\rangle$ for $N$-spin system, and the superposition state becomes
\begin{equation}    |\psi_{suppos}\rangle=\alpha|\alpha_1\alpha_2\dots\alpha_\text{N}\rangle+\beta|\beta_1\beta_2\dots\beta_\text{N}\rangle
\label{eq:b1}
\end{equation}
here $\alpha_i,\beta_i=\text{e},\text{g}$ with $i=1,2,\dots,\text{N}$ and $\alpha, \beta \in \mathbb{C}$. The magnetic moment~(\ref{eq3}) of the $j^\text{th}$ spin is calculating over the term
\begin{equation}
m_j=\langle\psi_{suppos}|
(\otimes_{i=1}^{j-1}\mathbb{I}_i)\otimes\hat{\sigma_j^z}\otimes(\otimes_{i=j+1}^N\mathbb{I}_i)|\psi_{suppos}\rangle,
\label{eq:b2}
\end{equation}
and substituting the~(\ref{eq:b1}) into the~(\ref{eq:b2}), we can easily separate the magnetic moment of the $j^{\text{th}}$ spin into two parts concerning the contribution from each superposed state, $|\alpha_1\alpha_2\dots\alpha_\text{N}\rangle$ and $|\beta_1\beta_2\dots\beta_\text{N}\rangle$ with the weights $\alpha$ and $\beta$, respectively.
\begin{equation}
    m_j=|\alpha|^2 m_j^\alpha+|\beta|^2m_j^\beta.
    \label{eq:b3}
\end{equation}
The superscripts $\alpha$ and $\beta$ denote the distinct contributions from each superposed state, $|\psi_1\rangle$ and $|\psi_2\rangle$. As we mentioned the probability conservation in the main text, $|\alpha|^2+|\beta|^2=1$. For simplify the notation, we will assume $|\alpha|^2\equiv p$ and $|\beta|^2\equiv(1-p)$. By utilizing the~(\ref{eq:b3}), we can get the EA SG-order parameter, $q_{\text{EA}}$~(\ref{eq2}) in terms of superposed states,
\begin{align}
q_{\text{EA}}&=\frac{1}{N}\sum_{i=1}^N (p~m_\text{i}^\alpha+(1-p)m_\text{i}^\beta)^2 \nonumber \\
&=\frac{1}{N}\sum_{i=1}^N (p^2~(m_\text{i}^\alpha)^2+(1-p)^2(m_\text{i}^\beta)^2+2p(1-p)m_\text{i}^\alpha m_\text{i}^\beta).
\label{eq:b4}    
\end{align}
According to the replica approach of Parisi~\cite{Parisi}, the EA order parameter measures the overlaps between two configurations. In our approach, the superposed states can be considered the replicas of the system. Once we get the system state in a superposition, $q_{\text{EA}}$ has already included the overlapping terms, $m_\text{i}^\alpha m_\text{i}^\beta$, beyond the self overlapping terms $(m_\text{i}^\alpha)^2$ and $(m_\text{i}^\beta)^2$.
\\
\newline
{\it{Equally weighted SSs}}. ($p=1/2$) The SG-order parameter~(\ref{eq:b4}) becomes for $\alpha=\beta$ (diagonal terms in Fig.~\ref{fig:4a}),
\begin{equation}
    q_{\text{EA}}(\alpha=\beta)= \frac{1}{N}\sum_{i=1}^N (m_\text{i}^\alpha)^2=1,
    \label{eq:b5}
\end{equation}
since $m_\text{i}^\alpha=\pm 1$. However, for the case $\alpha\neq\beta$ (off-diagonal terms in Fig.~\ref{fig:4a}),
\begin{equation}
 q_{\text{EA}}(\alpha\neq\beta)=\frac{1}{4N}\sum_{i=1}^N (m_\text{i}^\alpha+m_\text{i}^\beta)^2,
 \label{eq:b6}
\end{equation}
here ($m_\text{i}^\alpha+m_\text{i}^\beta)=\pm 1, 0$, and the square of the term gives the non-negative values ($m_\text{i}^\alpha+m_\text{i}^\beta)^2=1,0$. 

Assuming that the number of zero values are $n_C$ come from the single qubit superposition state ($C$), the remaining $(N-n_C)$ terms be equal to $1$, and the Eq.~(\ref{eq:b6}) become
\begin{equation}
 q_{\text{EA}}(\alpha\neq\beta)=\frac{N-n_C}{4N}=0.25-\frac{n_C}{4N},  
 \label{eq:b7}
\end{equation}
and it depends on the number of spins with zero magnetic moments ($n_C$) for finite system size $N$. We have found that an upper boundary of $q_{\text{EA}}\leq 0.25$ as $q_{\text{max}}=0.25$ for $N>n_C$. In terms of the probability of the $C$-qubits, $q_{\text{EA}}(\alpha\neq\beta)=q_{\text{max}}(1-p_C)$. 
In the lower and upper limit of the $p_C=n_C/N$;
\begin{itemize}
    \item for $n_C=0$ ($p_C=0$), $q_{\text{EA}}(\alpha\neq\beta)=q_{\text{max}}$,
    \vspace{3pt}
    \item for $n_C=N$ ($p_C=1$), $q_{\text{EA}}(\alpha\neq\beta)=0$,
    \item for $0<p_C<1$,~~~ $0<q_{\text{EA}}<q_{\text{max}}$.
\end{itemize}
The SG-order can be seen for the third condition, while the paramagnetic phase occurs in the second case. 

\section{Superposition Susceptibility}
\label{sec:A}

According to Parisi's mean-field theory, the susceptibility of a spin-glass in an applied field is obtained from the full $q$ SG-order parameter values (full Gibbs average) and interpreted as the field-cooled (FC) susceptibility~\cite{FCZFCSus}. Assuming that the system is restricted to be in one state, as in our case, the zero-field cooled susceptibility is obtained from a unique $q_{\text{EA}}$ given in Eq.~(\ref{eq11new}), $\chi_{ZFC}=\tilde\beta(1-q_{\text{EA}})$ in terms of $q_{\text{EA}}$. As we discussed before, the paramagnetic order occurs for $q_{\text{EA}}=0$ and the zero-field cooled susceptibility become
\begin{equation}
    \chi_{ZFC}=\tilde\beta,
\end{equation}
as it is convenient with Curie Law~\cite{curie}.

\bibliography{QSPSG_Reference}

\end{document}